\pdfoutput=1
\documentclass{aa}  
\usepackage{graphicx}
\usepackage{subfigure} 
\usepackage{txfonts}
\usepackage[figuresright]{rotating}
\usepackage{natbib}
\bibpunct{(}{)}{;}{a}{}{,}
\begin{document}

   \title{Spitzer observations of the N157B supernova remnant and its
     surroundings}

   \author{E. R. Micelotta, B. R. Brandl, F. P. Israel}

   \offprints{E. R. Micelotta          }

   \institute{Sterrewacht Leiden, Leiden University, P.O. Box 9513, 2300 RA 
          Leiden, The Netherlands\\
              \email{micelot@strw.leidenuniv.nl}
             }

   \date{Received 26 March 2008 / Accepted 16 February 2009}

  \abstract
   {}
{We study the LMC interstellar medium in the field of the nebula
  N157B, which contains a supernova remnant, an OB association,
  ionized gas, and high-density dusty filaments in close proximity.
  We investigate the relative importance of shock excitation by the
  SNR and photo-ionization by the OB stars, as well as possible
  interactions between the supernova remnant and its environment.  }
{We apply multiwavelength mapping and photometry, along with spatially
  resolved infrared spectroscopy, to identifying the nature of the
  ISM using new infrared data from the \textit{Spitzer} space observatory 
  and X-ray, optical, and radio data from the literature.  }
{The N157B SNR has no infrared counterpart. Infrared emission from
  the region is dominated by the compact blister-type HII region
  associated with 2MASS J05375027-6911071 and excited by an O8-O9
  star. This object is part of an extended infrared emission region
  that is associated with a molecular cloud.  We find only weak emission 
  from the shock-indicator [FeII], and both the excitation and the heating
  of the extended cloud are dominated by photo-ionization by the early
  O stars of LH~99.  }
{Any possible impact by the expanding SNR does not now affect the
  extended cloud of molecules and dust, despite the apparent overlap
  of SNR X-ray emission with infrared and H$\alpha$ emission from the
  cloud. This implies that the supernova progenitor cannot have been
  more massive than about 25 M$_{\odot}$.
   \keywords{supernova remnants -- N157B -- LMC -- H II regions -- dust -- 
             photoionization -- shock-excitation
            }}
   \authorrunning{E. R. Micelotta et al.}

   \titlerunning{Spitzer observations of LMC-N157B}
   \maketitle


\section{Introduction}

\begin{figure*}
     \centering
     \subfigure[]{
           \includegraphics[width=.42\textwidth]{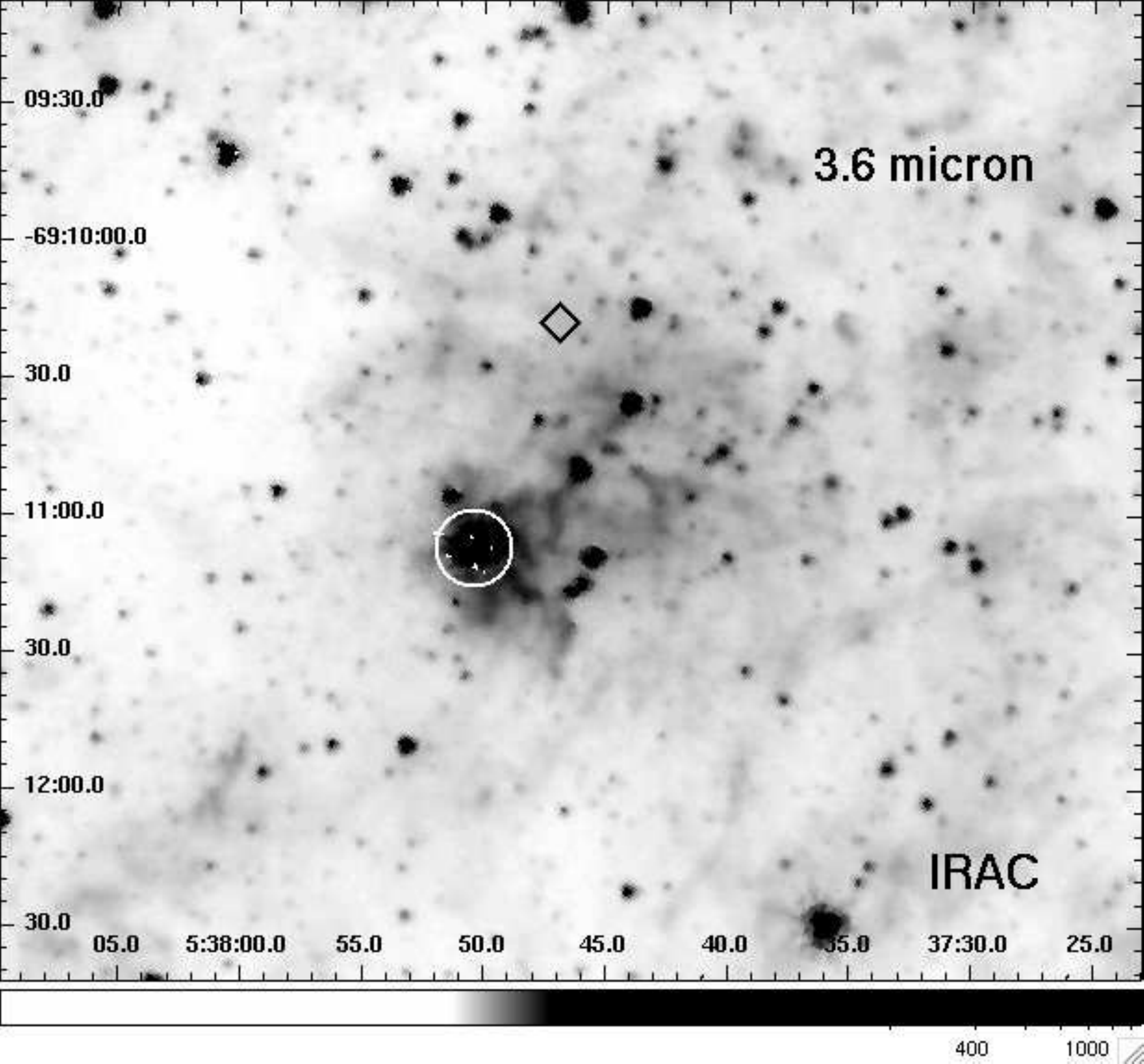}}
     \subfigure[]{
          \includegraphics[width=.42\textwidth]{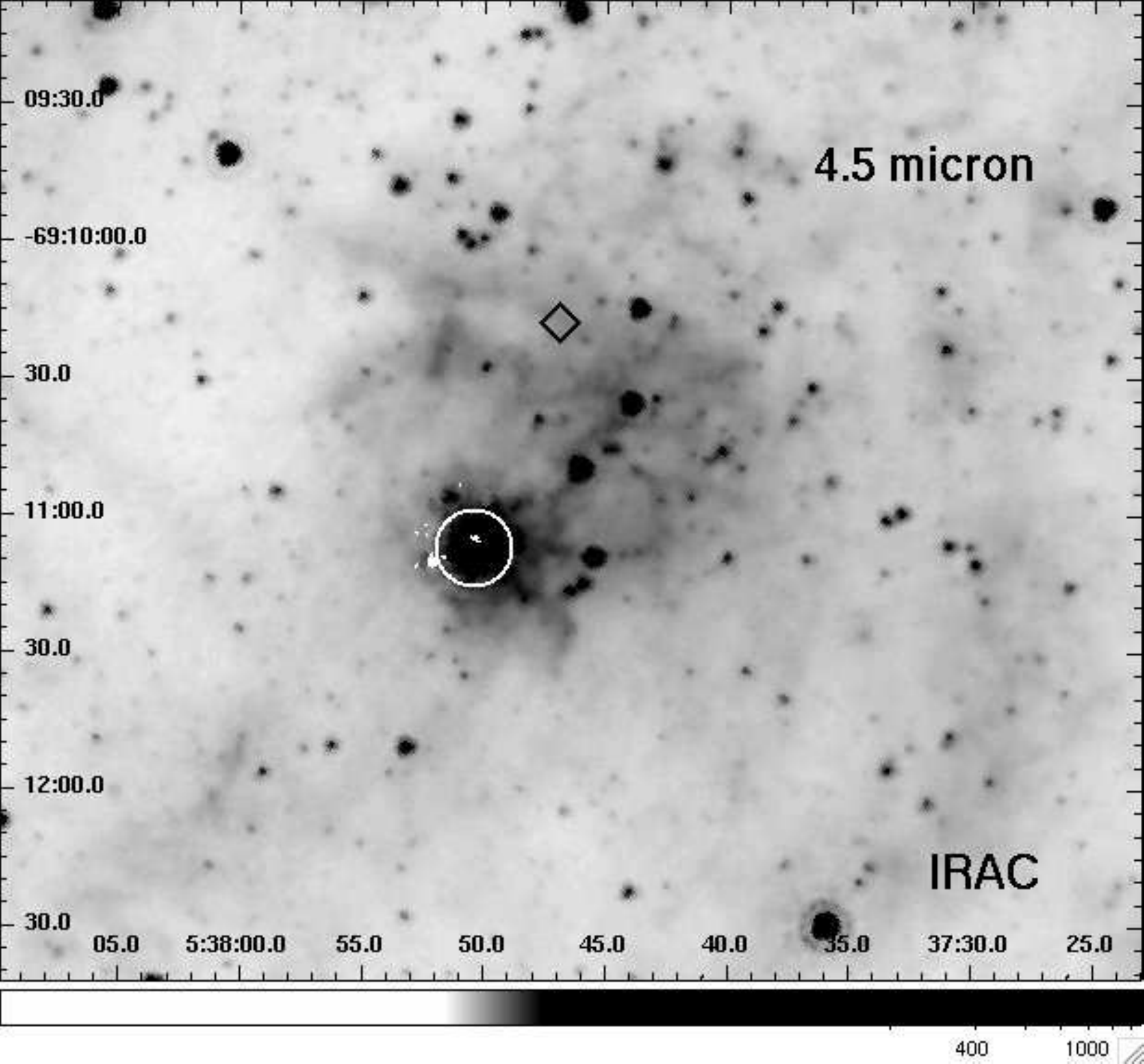}}\\
     \subfigure[]{
           \includegraphics[width=.42\textwidth]{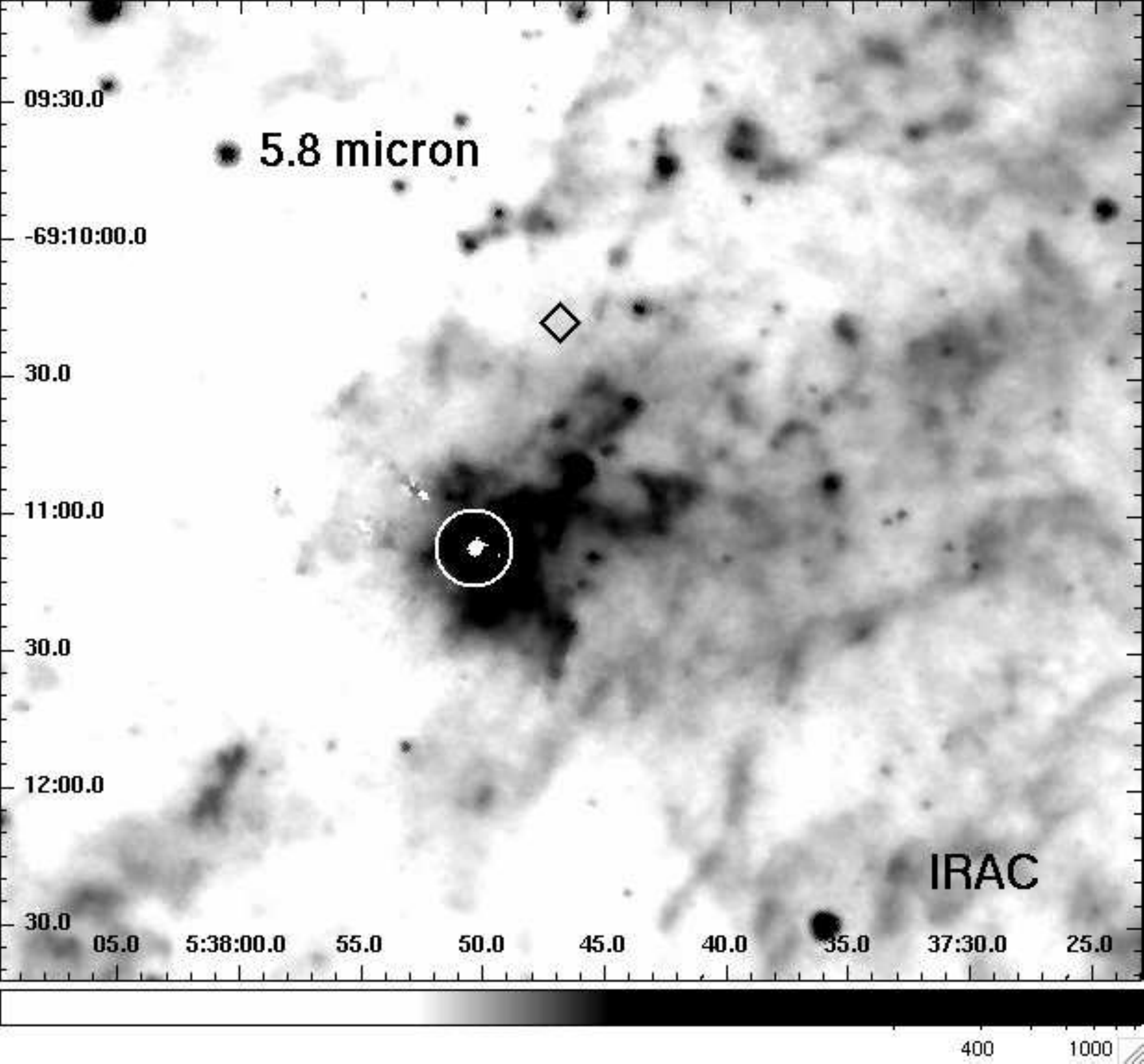}}
     \subfigure[]{
           \includegraphics[width=.42\textwidth]{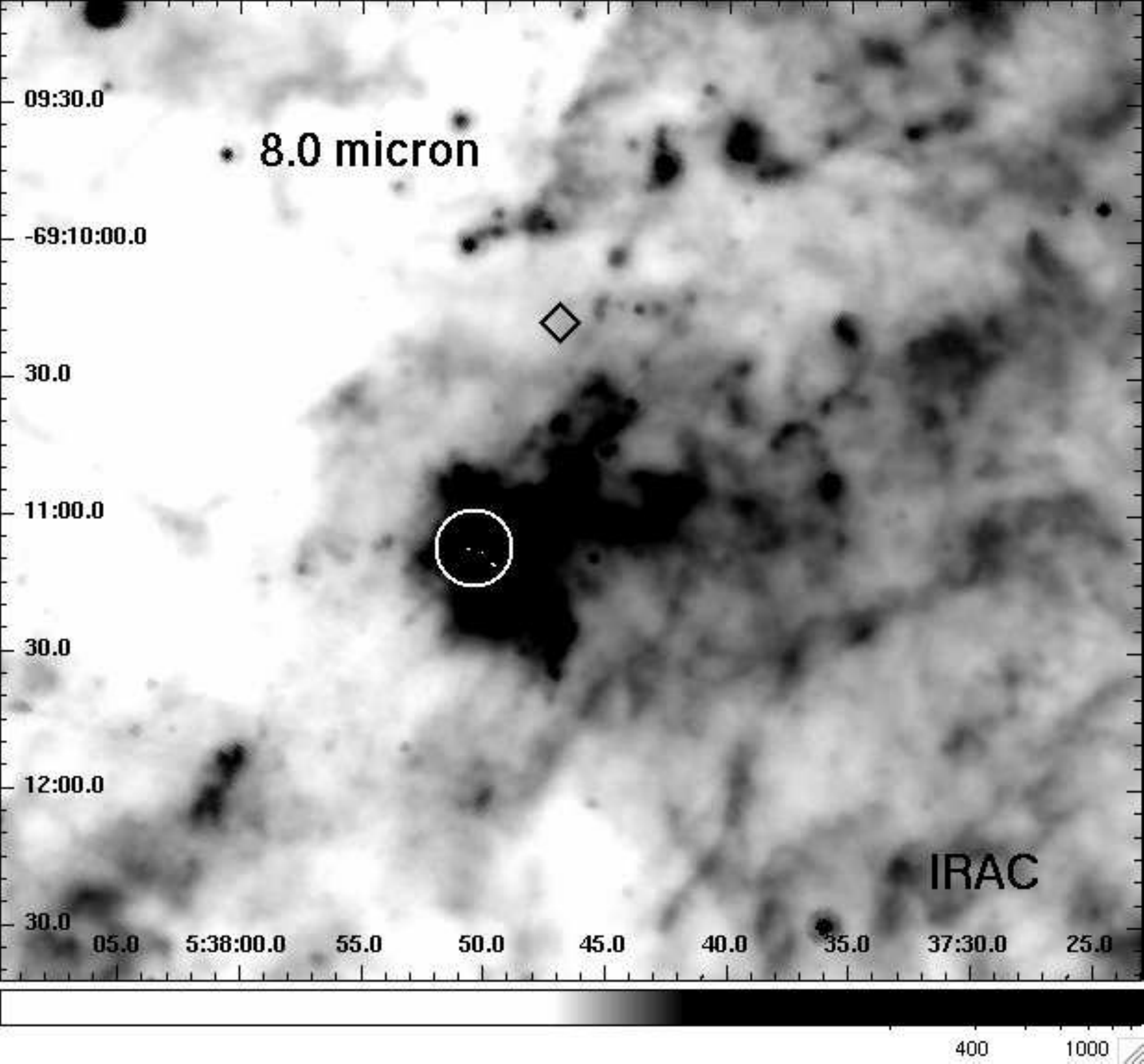}}\\
     \subfigure[]{
           \includegraphics[width=.42\textwidth]{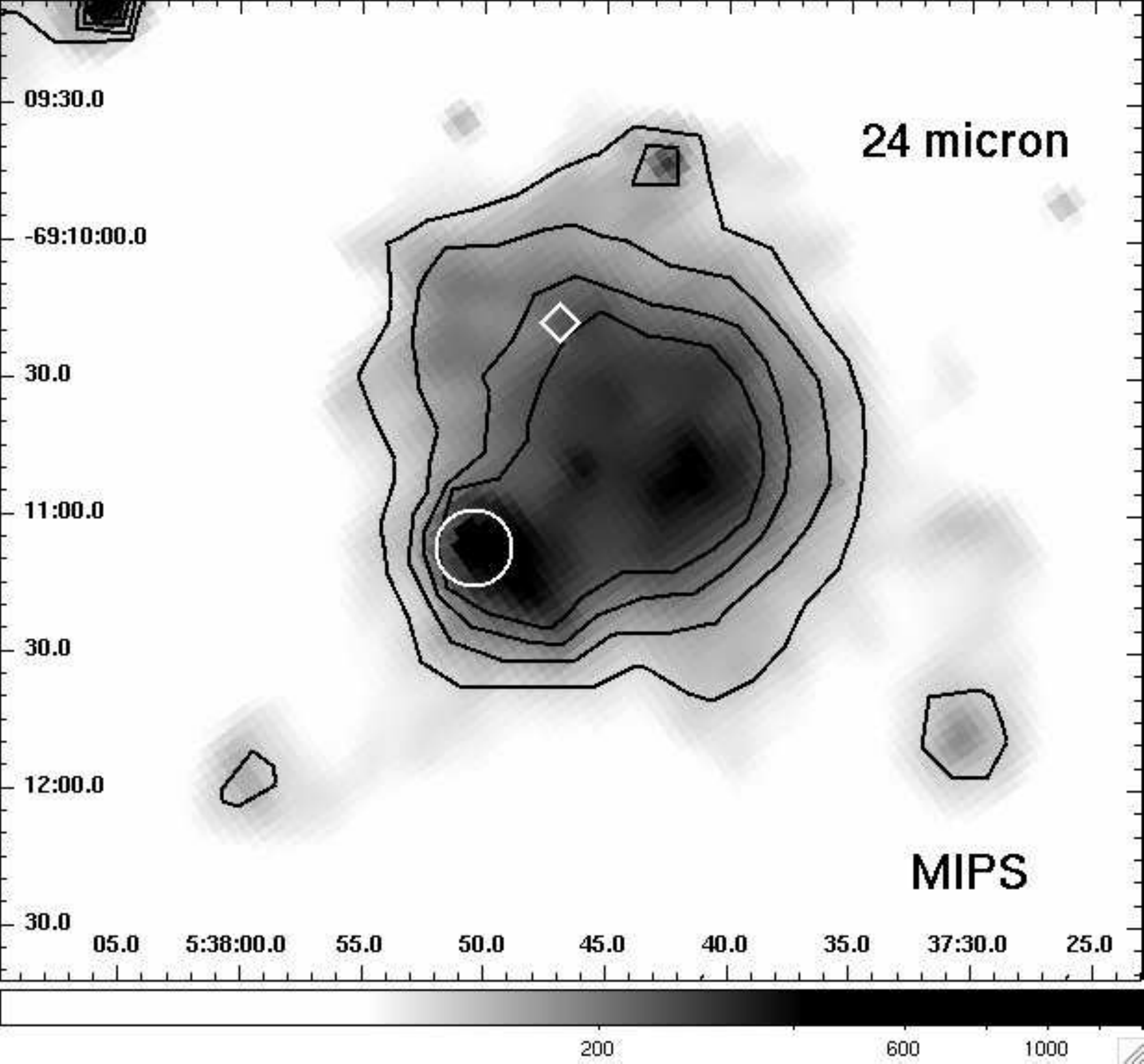}}
     \subfigure[]{
           \includegraphics[width=.42\textwidth]{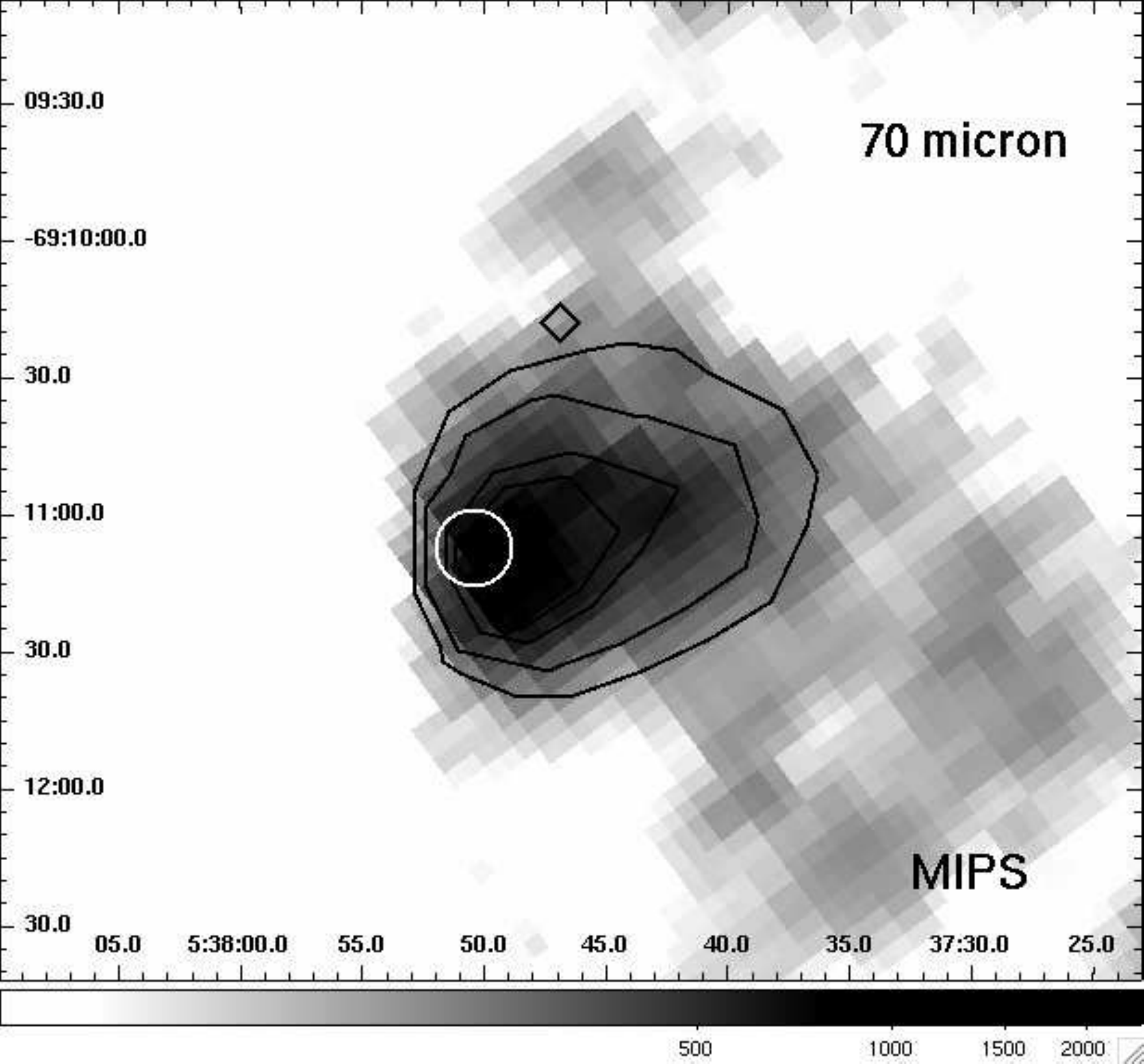}}\\
     \caption{IRAC and MIPS images of the N157B region,
              	calibrated in MJy/sr.
              The diamond marks the peak of the 
              supernova remnant X-ray emission \citep{sas2000} and the 
              circle marks the 2MASS compact source J05375027-6911071.
              The coordinates are (RA, dec) J2000.
              }
    \label{IRAC_MIPS_fig}
\end{figure*}

   \begin{figure*}
   \centering
   \includegraphics[width=18cm]{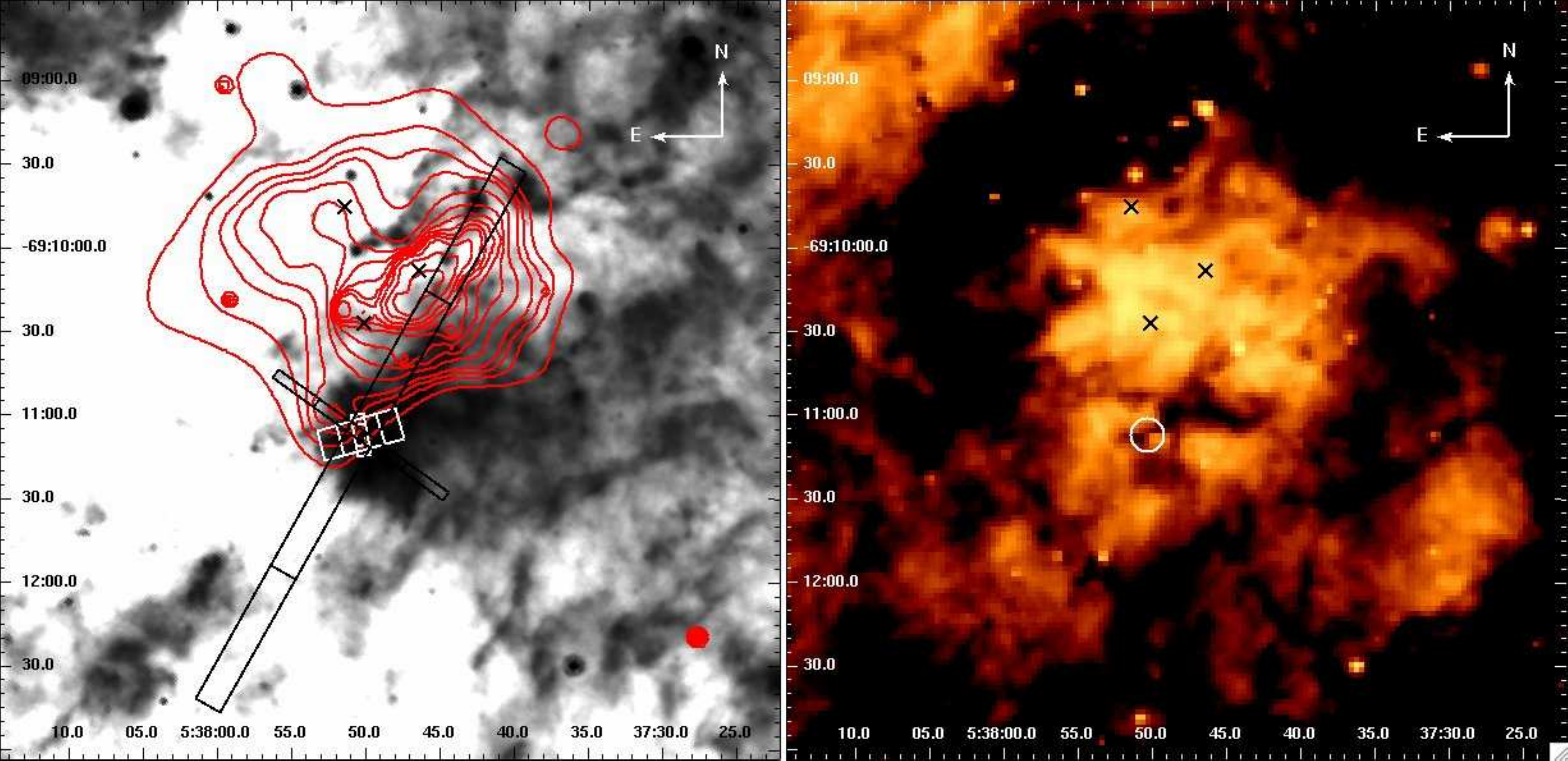}
    \caption{
      \textit{Left panel}: IRAC 8.0 $\mu$m map overlaid with
      Chandra ACIS-S X-ray intensity contours of N157B at 4, 5, 7, 8,
      12, 16, 20, 25, 30, 35, 50, 100, 500, 30000 $\times\,10^{-2}$
      counts sec$^{-1}$ arcsec$^{-2}$ (courtesy D. Wang). The
      rectangles mark the IRS slit locations: SL = east-west black,
      LL2 = north-south black, SH and LH are the two small white boxes
      in the center of the field.  The O3 stars ST~1-62, 1-71, 1-78 in
      LH~99 (\citep{schild92}) with projected positions inside the
      N157B X-ray contours are marked with an ``X''
      \citep{schild92}.  \textit{Right panel}: MCELS H$\alpha$ image
      of N157B \citep{smith98}, again with the O3 stars marked.
      Note the correspondence between the 8 $\mu$m IR emission and
      H$\alpha$ extinction. The compact object J05375027-6911071 is
      marked by a circle which also identifies the area over which we
      determined the H$\alpha$/[S II] ratio (see Sect. 4.1). 
      In both panels the coordinates are (RA, dec) J2000.
      }
   \label{all_fig}
   \end{figure*}

   \begin{figure}
   \centering
   \includegraphics[width=8.0cm]{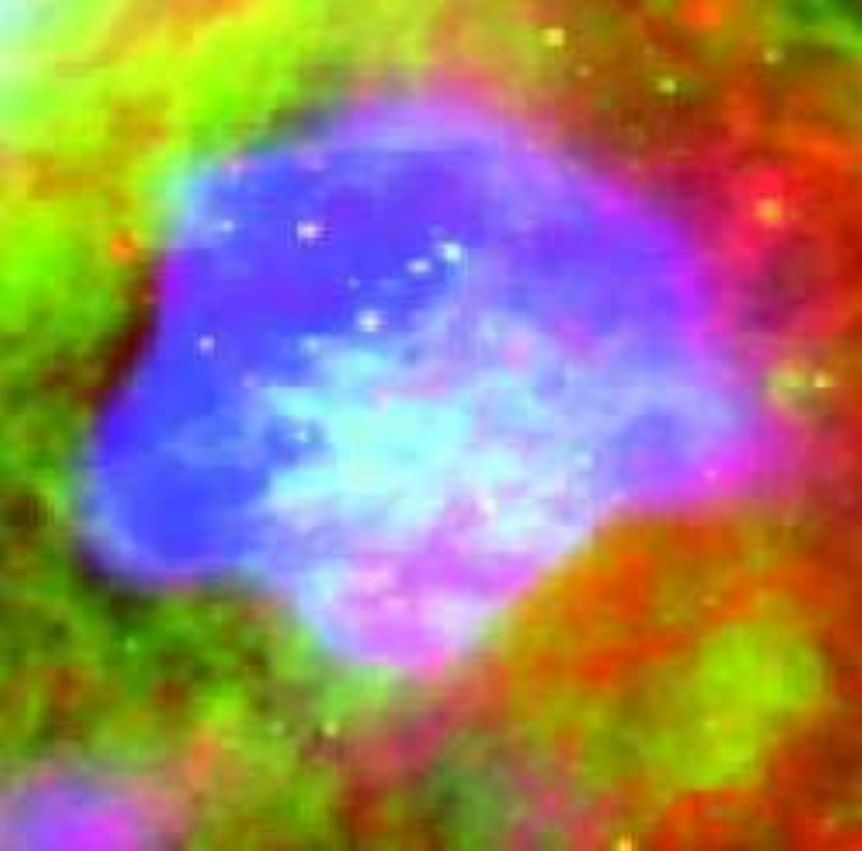}
      \caption{Composite image of the N157B region:
              red = Spitzer IRAC 8.0 $\mu$m, green = MCELS H$\alpha$, 
              blue = Chandra *ACIS* 0.9 - 2.3 keV. The blue area in the center
              shows X-rays from the northend parts of the remnant filling 
              a prominent hole in the 8.0 $\mu$m.
              	The contours of the optical and 
              infrared emission (see Fig. \ref{all_fig}) are recognisable 
              in the cyan and magenta areas respectively . 
              This picture is part of a large image of 30 Dor from
              Townsley et al. (2006). (\textit{Reproduced by permission of the AAS.})
              }
         \label{Ha_8um_X_fig}
   \end{figure}

The H$\alpha$-emitting region \object{N157~B} \citep{henize56} in the Large
Magellanic Cloud is on the southwestern fringe (projected linear
distance 90 pc from the center) of the major star-forming complex 
\object{30 Doradus}.  In a limited area of (projected) diameter 65 pc, the field
contains a supernova remnant, an OB association, HII emission regions,
bubbles of low-density hot gas, as well as the neutral material in 
dense clouds  (\cite{lazen00} and references therein).  The structure,
dynamics, and energy balance of this region are poorly understood.
There is even uncertainty about the nature of the emission observed at
different wavelengths.

The OB association \object{LH~99} \citep{luc70} is seen on the sky in the same
direction as N157B, and is often considered to be associated with it
because it has the same foreground extinction \citep{wang98}.  LH~99
contains a large number of (early) O stars \citep{schild92}, powerful
ionizing sources whose strong stellar winds are expected to produce
bubbles of low-density hot gas \citep{tow06} and to illuminate the
dusty filaments and clouds such as the nearby molecular cloud 30Dor-22
\citep{joh98}.

Much of the nebula N157~B can be identified with the Crab-like
supernova remnant (SNR) B0538-691.  Its embedded X-ray pulsar 
\object{PSR J0537-6910} suggests an age of about 5000 yr \citep{marsh98, wang98}.
The N157B radio counterpart MC~69 \citep{lemarne68, mcgee72} has a
spectral index $\alpha\,=\,-0.19$, $S_{\nu}\propto\nu^{\alpha}$ 
\citep{mills78, lazen00}.  Such a flat spectrum is characteristic
of thermal HII region emission, as well as of nonthermal Crab-type SNR
emission.  However, various observations strongly support the latter
interpretation: bright X-ray emission coincides with the radio source
\citep{long79}, the ratio of radio-to-optical intensity is quite high
\citep{dick94}, spectral line images \citep{danz81} reveal filamentary
structures with a high line ratio[SII]/H$\alpha\,\geq\,0.7$, the lines
of [OI], [FeII], and [FeIII] are strong, whereas the HeI $\lambda$4686
\AA{} line is weak.  The SNR has a peculiar one-sided morphology, does
not have a well-determined outer boundary, and exhibits bright
radio/X-ray core surrounded by an extended envelope of linear size of
30$\times$20 pc \citep{chu92, dick94, lazen00}, or even larger
\citep{tow06} making the remnant unusually large in both radio and
X-ray emission \citep{lazen00, wang98}.

In such a complex environment, shocks from SNRs and photons from
luminous stars compete as heating agents for the ISM.  The conditions
in and around N157B have therefore been a frequent subject of study.
Several authors, e.g. \citet{chu92, tow06, chen06} have concluded that
the expanding SNR is presently colliding with the molecular cloud to
its south.  The impact of a shock front on dense cloud material should
leave a signature detectable at mid-infrared (MIR) wavelengths
\citep{oliva99, reach00}.  Confirmation of the proposed
collision is of great interest, as it would identify an excellent
source for studying the physics of ongoing dense ISM processing by
shocks.

The use of infrared observations is particularly well-suited to trace
physical conditions in dusty environments because of their
insensitivity to extinction.  In the following, we combine our 
\textit{Spitzer} observations with literature data at a variety of
wavelengths to study the role of the different physical components
(SNR, HII region, OB association, dust clouds) and heating processes
(shock-excitation, photo-ionization) in shaping the N157B region.
Throughout, we adopt a distance of 50 kpc for the LMC \citep{west97},
so that 10 $''$ correspond to 2.4 pc.


\section{Observations and data processing}

We have observed N157B with the Infrared Array Camera (IRAC)
\citep{fazio04} and with the Infrared Spectrograph (IRS)
\citep{houck04} on board the Spitzer Space Telescope \citep{wern04},
as part of the IRS guaranteed-time program (PID 63, PI J. R. Houck).
The images from the Multiband Imaging Photometer for Spitzer (MIPS)
have been retrieved from the archive (MIPS 24 $\mu$m: program 3680, PI
K.J. Borkowski, MIPS 70 $\mu$m: program 20203, PI M. Meixner).

The IRAC data (Fig.\,\ref{IRAC_MIPS_fig}) were taken on 7 December 2003
and consist of mosaic images in
four channels at 3.6, 4.5, 5.8 and 8.0 $\mu$m.  The Basic Calibrated
Data (BCD) products from the Spitzer Science Center (SSC) pipeline
were used to construct the mosaic images.

The IRS spectra were taken on 26 May 2005 using the standard IRS
``Staring Mode'' Astronomical Observing Template (AOT).  The
observational setup is reported in Table \ref{setup_tab}.  The IRS
slit coverage is shown in Fig.\,\ref{all_fig} - left panel, outlined
with rectangles: SL goes from East to West, LL2 from North to South
and SH and LH are the two small rectangles in the center of the field.

%
\begin{table}
\begin{minipage}[t]{\columnwidth}
\caption{IRS observational setup. The IRS modules are SL: Short-Low, 
         LL2: Long-Low 2, SH: Short-High, LH: Long-High.} 
\label{setup_tab}      
\centering                          
\renewcommand{\footnoterule}{} 
\begin{tabular}{c c c c c}     
\hline\hline   
\noalign{\smallskip}              
Cycles  &  Duration$^{a}$  &  IRS   &  $\Delta\lambda$  &  Resolving Power  \\ 
number  &  (sec)     & module &  ($\mu$m)         &              \\
\noalign{\smallskip} 
\hline
\noalign{\smallskip}                        
$ $   2 & 14 & SL  & 5.2 $-$ 14.5  & 64 $-$ 128 \\ 
      2 & 14 & LL2 & 14.0 $-$ 21.3 & 64 $-$ 128 \\
      3 & 30 & SH  &  9.9 $-$ 19.6 & 600        \\
      3 & 60 & LH  & 18.7 $-$ 37.2 & 600        \\
\hline
\noalign{\smallskip}                                   
\end{tabular}
\end{minipage}
(a): Total integration time: SL + LL2 = 112 sec, SH + LH = 540 sec.\\
\end{table}
%

%
\begin{table}
\caption{Infrared flux densities.}
\begin{minipage}[t]{\columnwidth}
\label{phot_tab}
\centering
\begin{tabular}{c c c l l}     
\hline\hline
\noalign{\smallskip}
Origin    & $\lambda$ & Bandwidth & \multicolumn{2}{c}{Flux Density (Jy)} \\
          &($\mu$m)   & ($\mu$m)  & Compact   & Extended \\
          &           &           & Object    & Cloud \\
\noalign{\smallskip}
\hline
\noalign{\smallskip}
$ $ 2MASS &   1.24 &  0.16 &   0.0013 $\pm$ 0.0001  &        ---       \\
$ $       &   1.66 &  0.25 &   0.0022 $\pm$ 0.0002  &        ---       \\
$ $       &   2.16 &  0.26 &   0.017  $\pm$ 0.006   &        ---       \\
$ $ IRAC  &   3.6  &  0.75 &   0.13   $\pm$ 0.02    &   1.7 $\pm$ 0.3  \\
$ $       &   4.5  &  1.01 &   0.33   $\pm$ 0.05    &   3.2 $\pm$ 0.5  \\
$ $       &   5.8  &  1.42 &   0.66   $\pm$ 0.1     &   7.2 $\pm$ 1.1  \\
$ $       &   8.0  &  2.93 &   0.98   $\pm$ 0.2     &   19  $\pm$ 3.0  \\
$ $ MIPS  &  24    &  4.7  &   1.7    $\pm$ 0.3     &   38  $\pm$ 6.0  \\
$ $       &  70    & 19    &   3.6    $\pm$ 0.5     &   139 $\pm$ 21.  \\
$ $       & 160    & 35    &   2      $\pm$ 0.3     &   105 $\pm$ 16.  \\
$ $ IRAS  &  12    &  7.0  &            ---         &    3  $\pm$ 0.5  \\
$ $       &  25    & 11.2  &            ---         &   22  $\pm$ 3.0  \\
$ $       &  60    & 32.5  &            ---         &  124  $\pm$ 19.  \\
$ $       & 100    & 31.5  &            ---         &  312  $\pm$ 47.  \\
\noalign{\smallskip}
\hline
\end{tabular}
\end{minipage}
\label{irphot}
\end{table}
%

   \begin{figure}
    \centering
    \includegraphics[width=8.5cm]{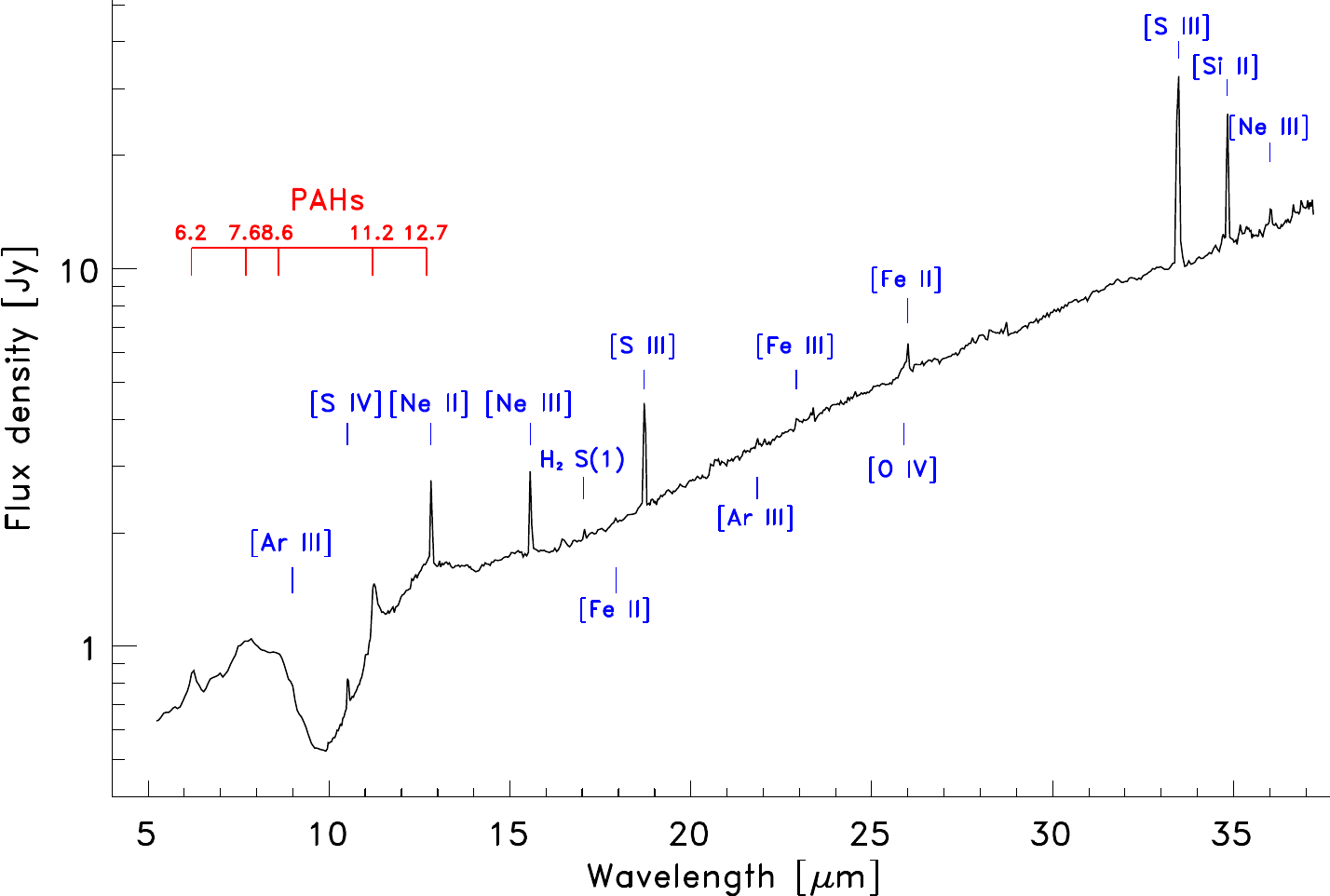}
    \caption{Combined low- and high-resolution IRS spectrum of the
      2MASS source J05375027-6911071.}
    \label{IRS_total}
   \end{figure}

Each cycle yielded two exposures at different nod positions along the
slit.  The data have been pre-processed by the SSC data reduction
pipeline version 12.0.2 (Spitzer Observer's Manual, chapter
7\footnote{http://ssc.spitzer.caltech.edu/documents/SOM}). The
two-dimensional BCD constituted the basis for further processing.
First we corrected the BCD frames for bad and rogue pixels with the
IRSCLEAN\footnote{available at
http://ssc.spitzer.caltech.edu/archanaly/contributed/}
tool. IRSCLEAN identifies bad and rogue pixels and replaces them with
the average of good nearest neighbors.  Rogue pixels are
single-detector pixels that show time-variable and abnormally high
flux values.  Next, we combined the frames from the same nod position
using the mean where two frames were available, and using the median
in case of three available frames.  For each integration at the same
position on the target, an off-source sky exposure is provided, so we
computed the mean/median sky (depending again on the number of
available frames) and subtracted it from the corresponding nod
position frame.

Further processing was done using the SMART package version 5.5
\citep{hig04} - a suite of IDL software tools developed for spectral
extraction and spectral analysis of IRS data.  The high-resolution
spectra were extracted using full slit extraction. The low-resolution
spectra were instead extracted using SMART's interactive column
extraction. The calibration is based on observations of standard stars
\citep{decin04}. The ends of each orders, where the noise increases
significantly, were manually clipped. To obtain a good match between
the low and high resolution modules, the SL + LL2 spectra were scaled
up by 10\%. Finally the resulting spectra from the two nod positions
were averaged to obtain the final spectrum shown in
Fig. \ref{IRS_total}.


\section{Results and analysis}

\subsection{IRAC and MIPS images}

The IRAC and MIPS images (Fig. \ref{IRAC_MIPS_fig}) of the N157B
region are dominated by emission from the dust associated with the
molecular cloud south of the X-ray emission region (peak marked by a
diamond).  Relatively faint in the 3.6 $\mu$m and 4.5 $\mu$m images, the
irregularly shaped diffuse emission is quite bright at 5.8 $\mu$m,
8.0 $\mu$m, and 24 $\mu$m where it appears more extended with a
quasi-circular shape.  In the 70 $\mu$m image, it no longer stands out
clearly as it suffers from confusion with the very extended
low-surface brightness emission characteristic of the whole 30 Doradus
region.  The cloud is listed as object no. 1448 in the IRAS catalogue
of LMC sources by \citet{schwe89} and shows up in the relatively
low-resolution IRAS maps as an extension of the main 30 Doradus IR
source.  Fig.\,\ref{all_fig} (left) and Fig.\,\ref{Ha_8um_X_fig} 
show that the contour of the 8 $\mu$m {\it emission} (sensitive to PAHs) 
from the cloud follows reasonably well 
the outline of {\it extinction} in the H$\alpha$ image (Fig.\,\ref{all_fig} - right).  
Much of the infrared-emitting dust must therefore be either in front of the
ionized gas, or embedded within it. From the MIPS 24 
$\mu$m, the size of the cloud is roughly $2'$ (28.8 pc). Its dimensions 
are thus similar to those of the SNR but the two are offset from one another, 
as shown by the superposition of the X-ray and 8 $\mu$m maps 
in Fig.\,\ref{all_fig}, left-hand panel.
Although there is considerable overlap between the X-ray and
H$\alpha$ emission regions (Fig.\,\ref{all_fig}, right-hand panel),
there is no trace of an IR counterpart to the X-ray SNR emission: the
two occur almost side-by-side.

The bright and compact object at RA (J2000) = 5h37m50.28s, DEC (J2000)
= -69$^{\circ}$11$'$07.1$''$ is located within the confines of the
infrared cloud.  In the IRAC images, it has a diameter of $\sim$3 pc
and it is the brightest source in the field in all IRAC bands.
Centimeter-wavelength radio maps by \citet{dick94} and \citet{lazen00}
reveal weak radio emission at its position. Although there is ISO SWS
and LWS spectroscopy of the extended cloud just described
\citep{verm02b}, the ISO apertures did not include this bright
object.  We have chosen this object, which is identical to the 2MASS
near-IR source \object{J05375027-6911071}, as the reference position for the
IRS slits (Fig.\,\ref{all_fig} - left panel).

We have collected infrared flux densities for both the compact object
and the whole dust cloud shown in Fig.\,\ref{IRAC_MIPS_fig} by
integrating the emission over circles with radii $6''$ and $57''$
centered on RA(5h37m50.3s), DEC(-69$^{\circ}$11$'$07$''$) and
RA(5h37m45.3s), DEC(-69$^{\circ}$10$'$47$''$), respectively.  The
results are listed in Table\,\ref{irphot}, which for convenience also
yields the relevant flux densities taken from the 2MASS on-line data
archive, and the IRAS database published by \citet{schwe89}. Although
all flux-densities in Table\,\ref{irphot} have very small formal
errors, a major uncertainty (easily a factor of two) arises in the
separation of the source from its surroundings.  As noted by
Schwering, this is clearly true for the IRAS flux densities, which are
very hard to separate from the overwhelming emission of the 30 Doradus
complex, but a glance at Fig.\,\ref{IRAC_MIPS_fig} shows that this
problem is not limited to the IRAS data, but also extends to the
Spitzer mapping of this complex region.

\subsection{IRS Spectroscopy of J05375027-6911071}


\begin{table}
\caption{Spitzer fine-structure lines observed in J05375027-6911071.}
\begin{minipage}[t]{\columnwidth}   
\label{lines_tab}      
\centering          
\begin{tabular}{l c c c r }
\hline\hline       
\noalign{\smallskip}
Line ID &  $\lambda_{rest}$    & EP$^{a}$  & Flux$^{b}$   & EW$^{a}$  \\
        & [$\mu$m] & [eV]    & [10$^{-20}$ Wcm$^{-2}$]    & [nm] \\ 
\noalign{\smallskip}
\hline                    
\noalign{\smallskip}
  \multicolumn{5}{c}{Clear detections}\\ 
\noalign{\smallskip}
\hline
\noalign{\smallskip}
$ $[S IV]   & 10.51  & 34.8 &  2.35  $\pm$ 0.26     &  10   \\
$ $[Ne II]  & 12.81  & 21.6 & 17.2   $\pm$ 4.0      &  80   \\
$ $[Ne III] & 15.56  & 41.0 &  8.4   $\pm$ 0.26     &  40   \\
$ $[S III]  & 18.71  & 23.3 & 13.3   $\pm$ 0.3      &  60   \\
$ $[Fe II]  & 25.99  &  7.9 &  3.32  $\pm$ 0.93     &  20   \\
$ $[S III]  & 33.48  & 23.3 & 55.6   $\pm$ 1.4      &  210  \\
$ $[Si II]  & 34.82  &  8.15& 20.7   $\pm$ 0.8      &  70   \\
$ $[Ne III] & 36.01  & 41.0 &  1.64  $\pm$ 0.52     &  5    \\
\noalign{\smallskip}
\hline
\noalign{\smallskip}
  \multicolumn{5}{c}{Marginal detections}\\ 
\noalign{\smallskip}
\hline
\noalign{\smallskip}                  
$ $[Ar III] & 8.99    & 27.6 &  1.20  $\pm$ 0.27     &  4.27   \\
$ $H$_{2}$ S(1) & 17.03 &  --&   0.8  $\pm$ 0.37     & 6.54  \\
$ $[Fe II]  & 17.93   &  7.9 &  0.17  $\pm$ 0.19     &  1.70   \\
$ $[Ar III] & 21.83   & 27.6 &  0.74  $\pm$ 0.20     &  4.00   \\
$ $[Fe III] & 22.92   & 16.2 &  1.47  $\pm$ 0.40     & 6.53  \\
\noalign{\smallskip}
\hline
\noalign{\smallskip}
\end{tabular}
\end{minipage}
Note. - Emission line properties obtained from single Gaussian fits in
the IRS spectrum (Fig. \ref{IRS_total}).\\ 
(a): EP = Excitation Potential, EW = Equivalent Width (observed).\\
(b): Quoted uncertainties are the errors from the line fit and do not include
calibration uncertainties.\\
\end{table}


\begin{table}
\begin{minipage}[t]{\columnwidth}
\caption{PAH dust emission features measured in J05375027-6911071.}
\label{PAHlowres_tab}      
\centering          
\renewcommand{\footnoterule}{} 
\begin{tabular}{r c c c}     
\hline\hline 
\noalign{\smallskip} 
$\lambda_{cent}^{a}$  &  
Flux$^{b}$ & 
EW$^{c}$ &
Relative$^{d}$ \\
$ $[$\mu$m] &        [10$^{-20}$ Wcm$^{-2}$] & [nm] & strength \\ 
\noalign{\smallskip} 
\hline \noalign
{\smallskip} 
 6.2 & 36.06 $\pm$ 0.58 & 34.1 & 1.00 \\ 
 7.7 & 53.00 $\pm$ 0.36 & 51.5 & 1.47 \\ 
 8.6 & 41.42 $\pm$ 0.54 & 40.3 & 1.15 \\ 
11.2 & 32.77 $\pm$ 0.64 & 36.7 & 0.90 \\ 
12.7 & 27.45 $\pm$ 0.36 & 26.7 & 0.76 \\ 
\noalign{\smallskip} 
\hline
\noalign{\smallskip}
\end{tabular}
\end{minipage}
(a): Central wavelength.\\
(b): The quoted uncertainties are the errors from the line fit and do not 
include the calibration uncertainties.\\
(c): EW = Equivalent Width (observed).\\
(d): Strength relative to 6.2 $\mu$m feature.\\
\end{table}      
%

   \begin{figure}
   \centering
   \includegraphics[width=8.5cm, height=8.5cm]{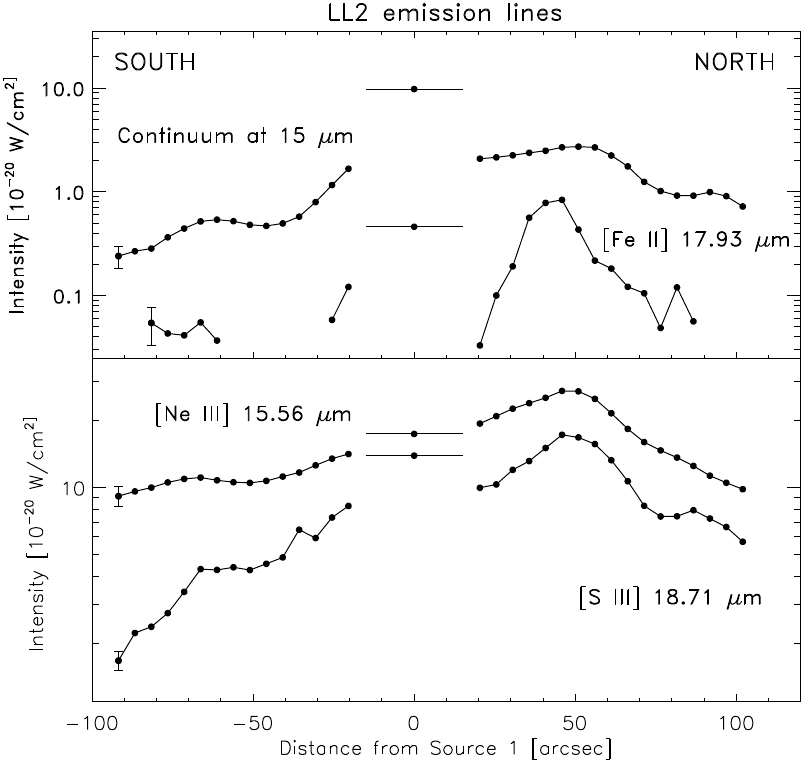}
   \caption{Spatial variation of continuum and emission line
     intensities along the low-resolution slit LL2,
     calculated with the program PAHFIT. The vertical bar on the left 
     mark indicates the representative error on fitted intensities.
     Spatial distances along the slit are measured from the position of
     J05375027-6911071 (0 arcsec).  The horizontal bar indicates the
     size of the five-pixel wide extraction window for
     J05375027-6911071; all other data points refer to three-pixel
     wide extraction windows. }
   \label{LL2_pos_orig}
    \end{figure}

   \begin{figure}
   \centering
    \includegraphics[width=8.5cm, height=6.5cm]{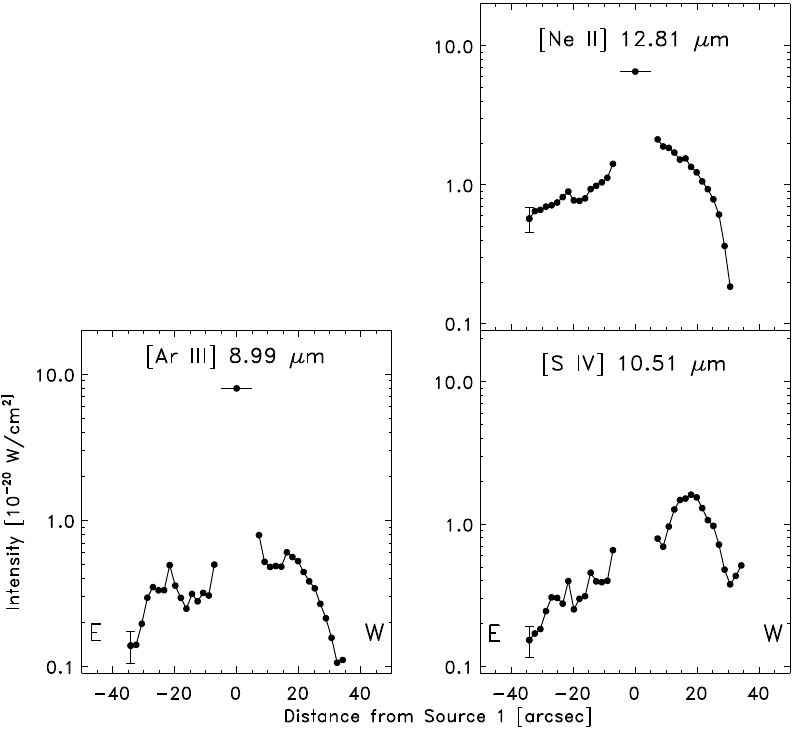}
    \caption{Spatial variation of emission line intensities along the
      low-resolution slit SL.  Otherwise as
      Fig. \ref{LL2_pos_orig}. Towards J05375027-6911071 the [S IV]
      line is undetectable because of strong silicate absorption.  }
              \label{SL_pos_lines_orig}
    \end{figure}

   \begin{figure}
   \centering
    \includegraphics[width=8.5cm, height=14.0cm]{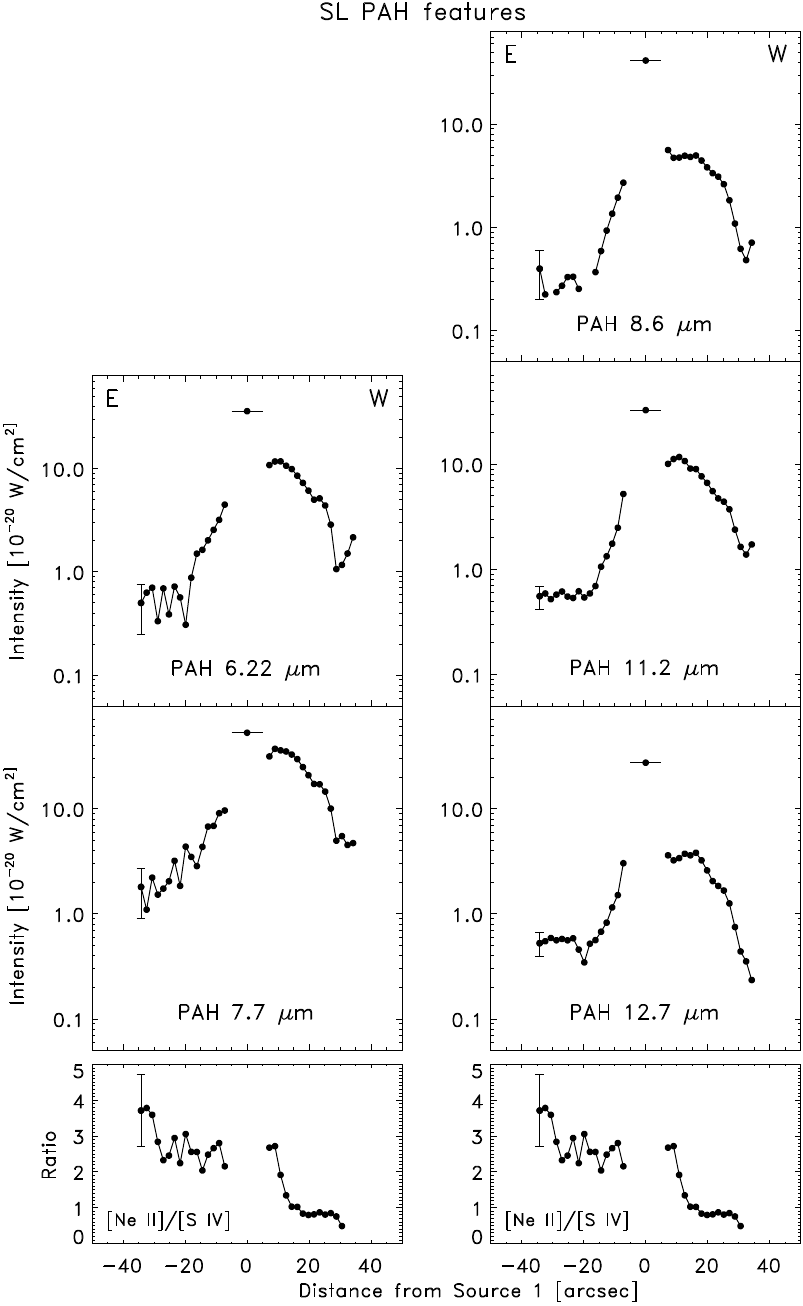}
    \caption{Spatial variation of PAH feature intensities (top panels)
      compared to the [Ne II]/[S IV] ratio (same plot reproduced in
      both bottom panels for clarity) along the low-resolution slit
      SL. The vertical bar on the left mark indicates the
      representative error on fitted intensities in the region between
      -35 and -5 arcsec. The oscillations are due to instabilities in
      the fit caused by weak signals. Otherwise as
      Fig. \ref{LL2_pos_orig}.  }
     \label{SL_pos_PAH_RATIO_orig}
    \end{figure}

The combined high- and low-resolution spectrum of the 2MASS source
J05375027-6911071 covers the wavelength range from 5 $\mu$m to 38 $\mu$m, 
and is shown in Fig. \ref{IRS_total}.  The (short-wavelength)
low-resolution part has been extracted from the {\it single slit
  segment} centered on 2MASS-J05375027-6911071 (see below), while the
high resolution part results from integration over the {\it full length}
covered by the SH and LH slits (see Fig. \ref{all_fig} - left-hand
panel). This area includes J05375027-6911071 but also some of its
fainter surroundings.  As a result, the spectrum in
Fig. \ref{IRS_total} accurately reflects the emission from
J05375027-6911071 shortwards of $\sim$20 $\mu$m.  However, at
wavelengths beyond 20 $\mu$m the spectrum includes a non-negligible
contribution from the surrounding diffuse emission.

Different ISM processes have left their mark on the spectrum: emission
from PAHs, absorption by silicates around 10 $\mu$m and in a broad band
between 15 $\mu$m and 22 $\mu$m, various fine-structure emission lines
and steeply increasing continuum emission from hot dust.  The dust PAH
features have been modelled with the program
PAHFIT\footnote{Available at
  http://turtle.as.arizona.edu/jdsmith/pahfit.php} \citep{smith07} and
the results are presented in Table \ref{PAHlowres_tab}.  PAHFIT is an
IDL tool for decomposing Spitzer IRS spectra, with special attention
to PAH emission features, and it is primarily designed for use with
Spitzer low-resolution IRS spectra.  The program is based on a model
consisting of starlight, thermal dust continuum with temperature from
35 to 300 K, resolved dust features and feature blends, prominent
emission lines and dust extinction dominated by the silicate
absorption bands at 9.7 $\mu$m and 15-22 $\mu$m.  PAHFIT uses Gaussian
profiles to recover the full strength of emission lines and Drude
profiles for dust emission features and blends.  The fine-structure
lines have been measured by single Gaussian fits within SMART, and the
results are listed in Table \ref{lines_tab}.  All these lines, except
the iron lines, are typically associated with ionized gas in
photon-dominated regions \citep[PDRs -  e.g.][]{mart02, tiel06}.

\subsection{IRS spectroscopy of the extended dust cloud}

As shown in Fig. \ref{all_fig}, the SH and LH slits cover only a
very limited part of the cloud, mainly the compact 2MASS object. The
two much longer low-resolution slits LL2 and SL sample a much larger
part of cloud, even extending into the cloud surroundings.  The slits
are almost perpendicular to one another and overlap at the position of
the compact object 2MASS J05375027-6911071.

In order to study the spatial variation of the spectral features along
the two slits, we sub-divided the region covered by each slit into 33
segments and extracted a spectrum from each segment.  We have chosen
overlapping extraction windows of three pixels width, moving them
along the slit in single-pixel steps. Thus, adjacent extractions are
not independent of one another as they overlap by two pixels. The
corresponding spectrum is therefore effectively a boxcar-smoothed
spectrum.  These particular choices resulted from tests performed on
the data to find the minimum extraction requirements needed to obtain
spectra free of sampling artifacts.  For the same reason, the region
containing the bright object J05375027-6911071 was treated slightly
differently.  Here, we applied an extraction window five pixels wide,
not overlapping with the adjacent pixels.  As a consequence, the size
of the extraction window for J05375027-6911071 exceeds the dimensions
of the intersection region common to both the SL and LL2 slits; it
corresponds to a 7 pc $\times$ 2.7 pc rectangle.  We used PAHFIT with
its default set of lines and continuum features to determine the
intensities of the spectral features at every slit position.  We have
plotted the intensities of fine-structure emission lines and dust
features thus extracted as a function of the position along the slit,
expressed in terms of the distance (positive and negative) from
J05375027-6911071 in Figs. \ref{LL2_pos_orig}, \ref{SL_pos_lines_orig}
and \ref{SL_pos_PAH_RATIO_orig}.

The SL spectra show PAH emission in the 6.2, 7.7, 8.6, 11.2 and
12.7 $\mu$m bands.  In determining the intensity of the latter, we
first removed the contribution from the [NeII] 12.8 $\mu$m line.  We
note that the PAH emission from the N157B region is remarkably weak
compared to that of other sources in 30 Doradus \citep{bra08}.


\section{Discussion}

\subsection{The nature of the compact object J05375027-6911071}

The flux densities in Table\,\ref{irphot} are not accurate enough to
warrant detailed SED fitting of J05375027-6911071. However, by
comparing various modified blackbody fits with straightforward
integration we find an integrated flux of $5-11\,\times\,10^{-13}$ W
m$^{-2}$ which implies a luminosity $L\,=\,3.8-8.6\,\times\,10^{4}$
L$_{\odot}$.  This luminosity corresponds to that of a B1-O9 star if
all stellar photons are converted to IR emission; it thus places a
lower limit on the spectral type of the exciting star(s).  Much of the
emission must arise from hot dust, with temperatures between 180 and
350 K. The exciting star must therefore be close to the dense neutral
material, and we would expect the interface between star and neutral
material to consist of very dense ionized gas. 

The \textit{Spitzer} spectrum of J05375027-6911071
(Fig. \ref{IRS_total}) contains several diagnostic fine-structure
lines.  The [SIII] lines at 18.7 $\mu$m and 33.5 $\mu$m arise from
different levels with the same excitation energy and their ratio
provides a measure of the electron density \citep[e.g.][]{tiel06}.  We
find a [SIII] 18.7 $\mu$m/ [SIII] 33. 5$\mu$m ratio of 0.24. Collisional
excitation models \citep{alex99} place this ratio in the low-density
limit, and indicate the presence of an ionized gas with
n$_{e}\,\approx\,100$ cm$^{-3}$.  Such a density is not uncommon for a
parsec-sized HII region \citep[cf. Fig.\,2 in][]{habing79}.  It is
also quite consistent with the weak radio emission (about 40 mJy at
$\lambda\lambda$3.5--13cm) in the maps published by \citet{lazen00} and
\citet{dick94}.  Assuming all radio emission in this direction to be
thermal and optically thin free-free emission, we calculate an
r.m.s. electron density $<n_{e}^{2}>^{0.5}\,=\,100-250$ cm$^{-3}$. It
is unlikely that the ionized gas and the hot dust have different
sources of excitation.  We must conclude that the bulk of the infrared
line and radio continuum emission arises from gas extended over a
volume much larger than occupied by the dense ionized gas surmised in
the previous paragraph.  As we have not resolved structure on scales
less than a few parsecs, this is quite feasible.

The ratios of lines of the same species but arising from different
ionization states reflect the degree of ionization and the hardness of
the stellar radiation field.  We have measured such pairs of neon and
sulphur lines and find ratios [SIV]/[SIII] = 0.18 and [NeIII]/[NeII] =
0.49.  In a sample of HII regions with different metallicities in the
Milky Way, the LMC, and the SMC these two ratios are tightly
correlated \citep[cf. Fig. 1 by][]{mart02}.  Our result fits this
correlation very well but the individual ratios are lower than those
in the (very) bright LMC HII regions which have [SIV]/[SIII] =
0.6--1.0, and [NeIII]/[NeII] = 1.4--6.3.  Not surprisingly, the gas in
J05375027-6911071 thus has a lower degree of ionization as is expected
from excitation by stars less hot than the ionizing stars in the
bright LMC HII regions.  The photoionization models by \citet{sch97}
(CoStar) and by \citet{paul01} show our line ratios to result from
stellar radiation fields with T$_{\rm eff}\,=\,33000-40000$ K
corresponding to spectral type O5--O9 \citep[cf.][]{martins02}.  Under
the same assumption of optically thin free-free radio emission, we
calculate from the maps by \citet{dick94} and \citet{lazen00} a minimum
required Lyman-continuum photon flux $N_{\rm L}\,=\,48.96$ s$^{-1}$. This
corresponds to the output of an O7.5--O9 star \citep{vacca96}.

Bright [FeII] 17.9 $\mu$m, [FeII] 26.0 $\mu$m and [SiII] 34.8 $\mu$m
lines trace the return of iron and silicon to the gas phase following
the destruction of dust grains by shocks \citep{oliva99, reach00}, 
but a strong [SiII] line is also frequently detected
in PDRs \citep{peet02}. In J05375027-6911071 [SiII] is one of the 
strongest detected lines, whereas the [FeII] lines are weak with 
respect to the others. The [SI] 25 $\mu$m line, also expected in 
shocked gas \citep{tiel06}, is missing but sulphur is detected as 
[SIII] and [SIV], a situation characteristic for photo-ionized gas.

The intensity of the optical lines [SII] $\lambda\lambda$6717, 6731
\AA{} and H$\alpha$ $\lambda$6563 \AA{} provides a reliable tool to
discriminate between shock-excited and photo-ionized plasma
\citep{long90}) and were, in fact, used to establish the nature of
N157B \citep{danz81}.  In SNRs, [SII] emission is usually stronger
than in photoionized HII regions, where sulphur is mostly doubly
ionized.  The value of the [S II]/H$\alpha$ ratio separating SNRs and
HII regions is about 0.4 \citep{dodo80, fesen85, long90}.  No optical
spectra are available for J05375027-6911071, but we may estimate the
[SII]/H$\alpha$ ratio from the MCELS line emission maps.  The lines
are close enough to assume that they suffer approximately the same
extinction.  We find a ratio [SII]/H$\alpha$ = 0.2, which is typical
for H~II regions \citep{long90}.  Note that in most of N157B, this
ratio exceeds $\sim$ 0.7 \citep{danz81}.

Finally, we note that the 9.7 $\mu$m silicate absorption feature is
very strong towards J05375027-6911071, with an optical depth
$\tau_{9.7}\,\approx\,1.1$.  This implies the presence of a large
column of neutral material in front of the ionized line emission
region.

Thus, the 2MASS source J05375027-6911071 contains a purely
photo-ionized gas of moderate density. The high column-density and
elevated temperature of the dust suggests a blister-type geometry
\citep{israel78} seen from the back, and a source of excitation located
close to the interface of the HII region with the obscuring dense
neutral material.  In that case, no more than about half of the
ionizing photons may escape, and we conclude from this and the local
radiation field hardness that the excitation of J05375027-6911071 is
caused by an obscured but no longer embedded O8 or O9 star.

\subsection{The northeast edge of the dust cloud}

The intensity variation of the spectral features along the rather
short SL slit is shown in Fig. \ref{SL_pos_lines_orig} and
\ref{SL_pos_PAH_RATIO_orig}.  The SL slit runs from northeast to
southwest.  It is potentially of interest as it cuts across the X-ray
SNR/IR dust cloud boundary (Fig. \ref{all_fig} left-hand panel).
Apart from J05375027-6911071, most of the emission in all lines comes
from the dust cloud, outside the X-ray contours.  The [S IV]
intensities peak around +15 arcsec (3.5 pc) southwest from
J05375027-6911071. Towards the object itself, because of the low 
resolution of the SL slit we could not separate the [SIV] emission 
from the silicate absorption at 9.7 $\mu$m, contrary to the high 
resolution spectrum where the [SIV] is well detected. Both [Ar III] 
and [Ne II] have intensity distributions strongly peaking on 
J05375027-6911071, and that of [ArIII] is almost symmetrical.  
The [NeII] distribution has an asymmetry similar to but less outspoken
than that of [SIV].  The ion with the highest ionization potential
(S[IV]) appears to trace the dust continuum best, but the lack of more
extensive spectral coverage makes it very hard to draw any solid
conclusions.

\begin{figure}
  \begin{minipage}{9cm}
    \includegraphics[width=8.5cm, height=6.0cm]{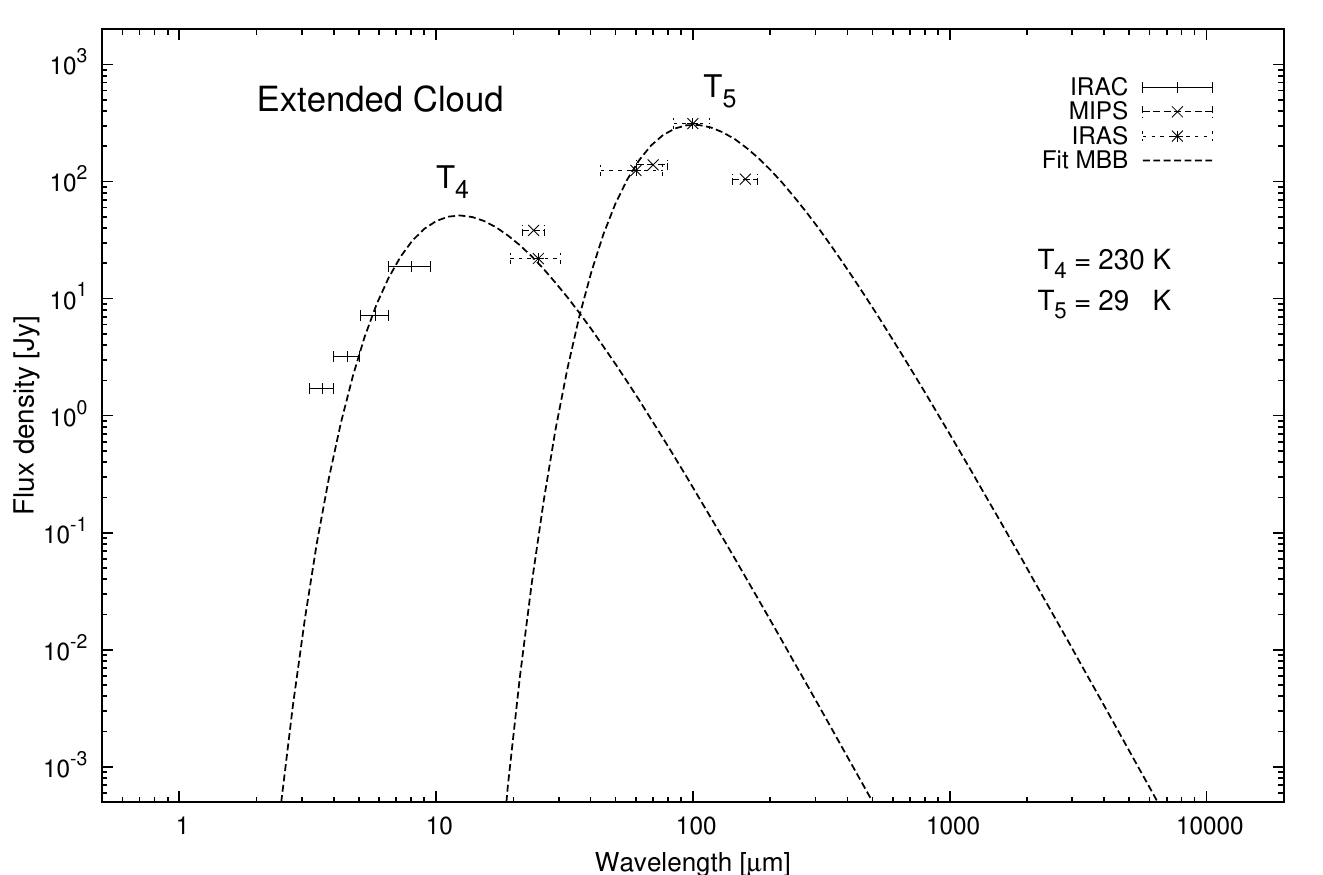}
  \end{minipage}
  \caption{Spectral energy distribution of infrared 
           emission from the extended dust cloud. The data points are 
           fitted with two modified black-body with $T\,=\,230$ and
           $T\,=\,29$ K. The horizontal bars indicate the width 
           of the IRAC \citep{fazio04}, MIPS \citep{rieke04} and 
           IRAS bands \citep{schwe89}. The 12 $\mu$m IRAS point has been
           removed because of the strong 10 $\mu$m silicate absorption
           of the compact source, which makes it impossible to separate
           out the extended cloud contribution.}
  \label{SED_fig}
\end{figure}

The various (weak) PAH features have symmetrical intensity
distributions, very similar to each other
(Fig. \ref{SL_pos_PAH_RATIO_orig}) and consistent with the 8.0 $\mu$m
image which shows the stronger emission in southwest of the X-ray
contours.  Fig.\,\ref{SL_pos_PAH_RATIO_orig} also shows the
[NeII]/[SIV] ratio. The excitation potentials of NeII
and SIV are 21.6 and 34.8 eV respectively. The ratio of the corresponding 
ionic lines [NeII]/[SIV] can be used as a tracer of the hardness of 
the interstellar radiation field (ISRF) in a similar way as the ratio
[NeII]/[NIII] \citep[see for exemple][]{giveon02}: a lower ratio 
corresponds to a harder ISRF. If PAHs were destroyed by FUV photons 
\citep[e.g.][]{madd06}, PAH intensities should be correlated with the 
[NeII]/[SIV] ratio. 

Inspection of Fig.\,\ref{SL_pos_PAH_RATIO_orig} reveals a clear trend: 
the PAH profiles peak where the [NeII]/[SIV] ratio dives ($+$20'' W) 
probably indicating the edge of an ionized region where PAHs are destroyed.
We refer for a more extensive and detailed discussion of
PAHs in the whole 30 Doradus region to \citet{ber08}.

\subsection{Conditions in the extended dust cloud}

Fig. \ref{LL2_pos_orig} depicts the spatial intensity variation of
spectral features along the significantly longer SE-NW LL2 slit.  The
slit extends from a region with weak IR emission and no X-rays to the
northwest, fully crossing the H$\alpha$ and X-ray emitting SNR
(Fig.\,\ref{all_fig} left-hand panel).  The compact source
J05375027-6911071 is brighter than its surroundings in the continuum,
and in the [SIII] and [FeII] lines, but not in the [NeIII] line.
Beyond this source, the distributions of [SIII], [NeII] and the
continuum are similar and peak about 50 arcsec (12 pc projected
distance) to its northwest.  The [FeII] from the diffuse cloud is
relatively sharply peaked.  Although uncertain, this peak may indicate
the presence of shocked gas at an SNR/dust cloud interface.  As the
Ne[III] and [SIII] intensity distributions peak closest to the early O
stars in LH~99 (see Fig.  \ref{all_fig} and Sec. 4.3), they seem to
reflect the radiative effects of these.

The ISO SWS and LWS instruments have provided spectroscopy from
differently-sized regions \citep{verm02a} roughly covering the O3
stars marked in Fig.\,\ref{all_fig}.  The LL2 slit misses the SWS
aperture completely, but does overlap with part of the LWS aperture.
The [NeIII] 15.6/36.0 $\mu$m, [SIII] 18.7/33.5 $\mu$m, and [OIII]
51.8/88.4 $\mu$m ratios (6.73, 0.40, 0.56 resp.) are all in the
low-density limit, indicating $n_{e}\,\leq\,100$ cm$^{-3}$.  The ISO
spectra also show [FeII] 26.0 $\mu$m and [SiII] 34.8 $\mu$m line
emission but not sufficiently dominant to be ascribed unequivocally to
shock excitation.  More importantly, the ISO data by \citet{verm02a}
imply ratios [SIV]/[SIII] = 0.25 and [NeIII]/[NeII] = 1.74.  These
values are also in excellent agreement with the relation established
by \citet{mart02}, and place the cloud somewhat closer to the bright
LMC nebulae.  Although there are complications briefly discussed by
\citet{mart02}, the ratios, taken at face value, suggest excitation by
O5-O6 stars rather than by O3 stars.

The spectral energy distribution of the extended infrared cloud is shown 
in Fig. \ref{SED_fig}. The emission peak is consistent with a modified 
black-body of temperature $T\,=\,29$ K 
($B(\nu, T) \times \nu^{\,\beta}$, with $\beta = 2$). 
Integration of the emission yields a flux of $1.1-1.5\,\times\,10^{-11}$ 
W m$^{-2}$ implying a luminosity $L\,=\,0.8-1.1\,\times\,10^{6}$ L$_{\odot}$. 
There is a second hot dust component, with a temperature $T\,\approx\,230$ K, 
and a luminosity (excluding emission from J05375027-6911071) 
$L\,=\,4-5\,\times\,10^{5}$ L$_{\odot}$. In the 12 $\mu$m
IRAS band the compact source has a strong silicate absorption at 10 $\mu$m,
which makes not possible the separation of the compact source and extended
cloud contributions. Before fitting the hot dust component we thus removed
the 12 $\mu$m IRAS point, for which the flux density is in fact 30 times lower 
than for the neighboring bands. 
Because of the very strong 
temperature dependence of dust emissivity, the mass in the hot dust 
component represents only a minute fraction of the total mass. The 
three O3 stars dominating the LH~99 population are projected onto 
the northeastern edge of the infrared cloud as depicted at 24 $\mu$m 
in Fig.\,\ref{IRAC_MIPS_fig}.  To their southwest,
i.e. closer to the brightest part of the nebula, there are at least 2
O5 stars, 5 O6 stars, and 5 O7 stars \citep[cf.][]{schild92}. Using the
luminosity scale by \citet{vacca96}, we find that these stars add
another $L$(O5-O7) = $4.1\,\times\,10^{6}$ L$_{\odot}$ to the
luminosity $L$(O3) = $3.6\,\times\,10^{6}$ L$_{\odot}$ already
provided by the brightest stars.  The hardness of the resulting mean
radiation field should therefore not be very different from the one
estimated above.

Emission from the radio source appears to avoid the dark cloud \citep[see
Fig.\,4 in][]{dick94}. This is not expected if the dust cloud is
only a foreground object, and it therefore indicates physical contact
between the radio (SNR) and infrared (molecular cloud) sources
respectively.  However, the infrared cloud reradiates at most 15$\%$
of the total photon output of the association members, requiring it to
cover only about a steradian as seen from the average O star.  LH~99
should thus be deemed to be quite capable of providing the energy to
power the infrared cloud, and there seems to be no need to invoke
energy inputs provided by an SNR impact.

\citet{chu92} have suggested that no more than about 20$\%$ of the
radio emission from MC~69 is thermal in origin, so that
$S_{\rm 5GHz}(th)\,\leq\,0.45$ Jy \citep[cf.][]{lazen00}. This requires a
maximum Lyman continuum photon flux $N_{\rm L}\,\leq\,1.0\,\times\,10^{50}$
s$^{-1}$.  The three somewhat excentric O3 stars provide a combined
flux $N_{\rm L}\,=\,2.5\,\times\,10^{50}$ s$^{-1}$, and the more embedded
later-type O stars an almost equal $N_{\rm L}\,=\,2.4\,\times\,10^{50}$
s$^{-1}$. Thus, the UV photon flux likewise should be sufficient to
provide for the ionization of the HII nebula associated with the SNR
and dust cloud.

\subsection{The SNR revisited}

The actual location of the cloud with respect to the remnant is still
poorly determined. The data presented here have revealed no evidence
for shocked material, although the dense cloud of dust and molecules
does appear to be quite close to the SNR.  Is the absence of such
impact (shock) signatures consistent with the dynamical evolution of
the remnant?  We assume that the remnant expansion is in the Sedov
phase \citep{wang98}, so that the temporal evolution of the
spherical blast wave with radius $R_{S}$, produced by the supernova
explosion into a medium of uniform density $\rho_{0}$, is described by
\citep{sedov59, mac79}:\\
$R_{\rm S} = 1.15\,(E_{0}/\rho_{0})^{1/5}\,t^{2/5}\,\approx 13\,(E_{51}/n_{0})^{1/5}\,t_{4}^{2/5}\;\;\ $ pc. \\
We assume an initial energy $E_{51}$ = $E_{0}/(10^{51}$ erg s$^{-1}$)
= 1, a remnant age $t_{4}$ = $t/(10^{4}$ yr) = 0.5 \citep{marsh98},
and a remnant radius $R_{\rm S}\,\simeq\,15$ pc \citep{chu92, dick94,
  lazen00, chen06}. Thus, the present size of the supernova remnant
requires an initial density of no more than $n_{0}\,=\,0.12$ cm$^{-3}$.
This rules out the possibility of N157B having expanded into an
interstellar gas cloud of any significant density, and confirms the
notion that expansion occurred inside a windblown cavity cleared out
by the supernova progenitor \citep{chu92}.  In their X-ray study of 30
Doradus and N157B, \citet{tow06} present a composite
X-ray/H$\alpha$/mid-IR image (their Fig.\,14) that indeed shows a
shell of ionized gas and warm dust surrounding the SNR.  To verify
whether a suitable low-density cavity could have been generated by the
supernova progenitor, we calculate its expected size by 
\citep[and references therein]{mac79}:\\
$   R_{\rm S} \approx 27\,L_{36}^{1/5}\,n_{0}^{-1/5}\,t_{6}^{3/5}\;\; $pc \\
where $R_{\rm S}$ is the cavity radius, $L_{36}$ the constant wind
luminosity in units of 10$^{36}$ erg s$^{-1}$, $n_{0}$ the
ambient gas particle density in cm$^{-3}$, and $t_{6}$ the time
that the wind has been blowing in millions of years. The wind
luminosity $L_{36}$ is given by:
$ L_{36} \approx 0.3 (M_{\ast}/20 M_{\odot})^{2.3}\;\; $ erg s$^{-1}$
We adopt a $M_{\ast}\,=\,25$ M$_{\odot}$ (corresponding to an O8--O9
star), intermediate between the values predicted by the \citet{thiel96}, 
\citet{heger03} and \citet{woos95} models so
that $L_{36}\,=\,0.5$.  The relevant time $t_{6}\,=\,3.2$ is the
entire lifetime of the progenitor, derived from e.g. \citet{kart95}:
$ t \sim (M_{\ast}/M_{\odot})^{-2.5} \times 10^{10}\;\; $yr.
The observed radius $R_{S}$ of about 15 pc implies an average density
of $n_{0}\,=\,310$ cm$^{-3}$ for the ambient medium into which the
windblown shell has expanded.  Such a value is compatible with the
densities of a few hundred per cc that we estimated in the previous
section for the clouds in the region.  Even the larger radius
$R_{S}\,\approx\,22$ pc gleaned from \citet{tow06} can be
accommodated as it still yields $n_{0}\,=\,50$ cm$^{-3}$.  This
situation changes rapidly, however, if we consider a more massive
supernova progenitor.  For instance, for $M_{\ast}\,=\,50$ M$_{\odot}$
and $R_{S}\,=\,15$ pc we have $L_{36}\,=\,2.5$ and $t_{6}\,=\,0.6$,
and find an average ambient density of only $n_{0}\,=\,9$ cm$^{-3}$.

Thus, we conclude that the lack of clear indications for an SNR impact
on the dust cloud seen in the IRAC and MIPS images is consistent with
the supernova explosion of a moderately massive star.  Only now would
the SNR expanding in the cavity begin to overtake the windblown shell
produced by the star over its lifetime.  The apparent lack of shocks
is not easily reconcilable with a supernova progenitor as massive as
the presently most luminous members of the LH~99 association.  This
confirms an independent conclusion by \citet{chen06} that the
progenitor should have been in the narrow mass range $M\,=\,20-25$
M$_{\odot}$.


\section{Conclusions}
From an analysis of Spitzer photometric mapping and spectroscopy of
the SNR-dominated region N157B in the LMC, we find that:

 \begin{itemize}
  \item There is no evidence of an infrared counterpart to the
    supernova remnant in the IRAC and MIPS images.
  \item The infrared emission is dominated by a cloud of dust and
    molecular gas adjacent to the remnant, containing the compact
    2MASS source J05375027-6911071. 
  \item The object J05375027-6911071 has a diameter of about 3 pc, an
    electron-density of 100-250 cm$^{-3}$, and is photo-ionized by an
    O8--O9 star. It is probably an open HII blister structure, seen
    from the back.
  \item In spite of the projected overlap between the SNR X-ray
    emission and the infrared cloud, there is at best very marginal
    evidence of shocked gas, while almost all data suggest
    photo-ionization and photon-heating to be the mechanisms
    dominating the infrared cloud.
  \item The extended dust cloud is associated with ionized emission of
    a density of typically 100 cm$^{-3}$, presumably at the edges of a
    denser molecular cloud.  As the extended dust reradiates only
    about 10 per cent of the luminosity of the 15 brightest and nearby
    O stars in the LH~99 OB association, these stars are sufficient to
    explain the heating of the dust cloud.
  \item The absence of clear evidence of shocks implies that at
    present the molecular/dust cloud is not significantly impacted by
    the remnant.  This suggests that the supernova progenitor was a
    moderately massive star of mass $M\,\approx\,25$ M$_{\odot}$.
  \end{itemize}

\begin{acknowledgements}
We gratefully acknowledge D. Wang, J. Dickel, and R. Gruendl for
providing Chandra data, radio continuum observations, and information
on the N157B stellar population, and we would like to thank the referee
for careful reading and useful comments. E.R.M. thanks A. Jones and
A. Tielens for support and useful discussions and acknowledges
financial support by the EARA Training Network (EU grant
MEST-CT-2004-504604). This work is based in part on observations made
with the \textit{Spitzer Space Telescope}, which is operated by the
Jet Propulsion Laboratory, California Institute of Technology, under
NASA contract 1047.
\end{acknowledgements}

\end{document}